# Delayed burst of particles from Ar matrices doped with $CH_4$ and radical-radical interaction

E. Savchenko[1], I. Khyzhniy[1], S. Uyutnov[1], M. Bludov[1], and V. Bondybey[2]

[1]B. Verkin Institute for Low Temperature Physics & Engineering NASU, Kharkiv 61103, Ukraine,

[2]Lehrstuhl für Physikalische Chemie II TUM, Garching b. Munich 85747, Germany

**Abstract**

The delayed explosive desorption of particles from Ar matrices doped with methane stimulated by an electron irradiation was studied in the extended concentration range from 1% to 10% using emission spectroscopy methods. Registration of the cathodoluminescence (CL) spectra of $CH_4$-containig Ar matrices revealed following products of radiation-induced methane transformation: H, CH and C. Three series of experiments were performed with different irradiation and heating modes: (i) continuous irradiation at low temperature with accumulation of radiolysis products, (ii) heating of the pre-irradiated films with measurements of relaxation emissions, specifically thermally stimulated exoelectron emission (TSEE) and (iii) external heating of the pre-irradiated films under beam. These measurments of the, so called, nonstationary luminescenve (NsL) at the selected wavelenghs and nonstationary desorption (NsD) in combination with TSEE registration provided an information on reactions of charged and neutral species of interest. Taking into account that the CH radical can be considered as a signature of the $CH_3$ species [27] we obtained the information on a role of H atoms and $CH_3$ radicals in stimulation of explosive delayed desorption of particles from Ar matrices doped with methane. Two bursts of long-period self-oscillations together with short-period ones with a limited number of periods were observeed during stationary irradiation of all methane-doped Ar matrices at low temperatures. Processes leading to the self-oscillations of particle yield are discussed.

**1 Introduction**

The ongoing interest in research of radiation effects in solid $CH_4$ and $CH_4$–containing ices is driven by its significant role in astrophysics, technology of cryogenic moderators as well as radiation physics and chemistry. The phenomenon of delayed explosive desorption of solid methane was observed during irradiation with neutron flux [1, 2], ions [3, 4] and electrons of subthreshold energy [5]. This kind of desorption was also found when a solid mixture of $CH_4$ and CO was exposed to fast electrons [6]. A certain delay in the explosion of particles relative to the onset of irradiation indicates that a necessary condition for initiating this type of desorption is the accumulation of radiolysis products and their subsequent reactions. Long-period bursts of delayed desorption from solid methane moderator, so-called "burping", were first detected by John M. Carpenter upon irradiation with a neutron flux [1]. During prolonged exposure to irradiation, periodic rises of temperature of several tens of degrees Kelvin were observed at intervals of about 24 hours at T=14 K and the destructive one took place after 334 h at 9 K. Such periodic bursts of temperature represent the case of self-oscillations [7]. The spontaneous release of energy, which ultimately leads to the destruction of the moderator's container, attracted considerable attention and stimulated further research [8–12]. The phenomenon was explained by the accumulation of radicals to some critical concentration upon a neutron flux followed by their recombination with a stored energy release. It should be noted that the electronically excited states of $CH_4$ molecule are prone to dissociation into fragments. The most effective fragmentation channel for a methane molecule $CH_4$ under the influence of high energy irradiation is the formation of a methyl radical $CH_3$ and hydrogen atom H: $CH_4 + \Delta E \rightarrow CH_3 + H$. Calculations of the processes of recombination of defects (radiolysis products) created by irradiation with heat release made in the approximation of a generalized type of defects [1] of the same type showed the occurrence of temperature and defects density self-oscillations in the system. Self-oscillations are determined by the activation nature of the defects' recombination rate caused by the temperature dependent diffusion of radicals. To describe the experiment, the activation energy of defect diffusion in methane was taken equal to about 150 K in Ref. [1]. More complex scenario was suggested in [2, 10]. The self-oscillations in methane under neutron irradiation were considered for the case of two types of defects (radiolysis products) created by irradiation, which were supposed to be:

$$H + H \rightarrow H_2 + \Delta E_1$$

$$CH_3 + CH_3 \rightarrow C_2H_6 + \Delta E_2$$

where $\Delta E_1 = 218$ kJ/mol, $\Delta E_2 = 368$ kJ/mol. The corresponding thermal activation energies for defect diffusion were taken as $\Delta E_1/k_B = 108.6$ K and $\Delta E_2/k_B = 235$ K [10]. Results of these calculations based on the theory of thermal explosion, demonstrated a good agreement with the test data obtained using methane moderators developed at the IPNS neutron source at Argonne National Laboratory as well as the data acquired during commissioning of the ISIS Target Station 2 solid methane moderator demonstrated long-period self-oscillations. Irradiation with ions, electrons and photons was used for laboratory simulation of conditions in outer space. Because of the significant content of $CH_4$ in the planetary and interstellar medium [13, 14], the radiative behavior of solid $CH_4$ and its mixtures has become the subject of a large number of studies, incl. [3, 4, 6, 15-23]. A fundamental problem in astrophysics is the existence of complex organic molecules in the gas phase of cold molecular clouds at temperatures as low as 10-20 K. One of the hypotheses suggested for explanation of this fact is based on explosive delayed desorption mechanisms driven by a rapid ejection of molecules stimulated by the release of "chemical energy" from the reactions of radicals stored in the ice mantles (see, for example, [6] and references therein). The detailed spectroscopic tracing of such radical reactions in mixture of $CH_4$ and CO exposed to superthreshold energy (5 keV) electrons was presented in [6]. During the irradiation, the chemical evolution of the ices was monitored online via Fourier transforminfrared (FTIR) spectroscopy. The molecules in the gas phase were detected using a quadrupole mass spectrometer. It was found that after some delay the concentrations of $CH_3$ and HCO radicals decreased sharply and this decrease was accompanied by a simultaneous increase in the concentrations of $C_2H_6$ and C=O group as well as an explosive rise in the chamber pressure. The role of such kind reactions in the cryovolcanism of comets was discussed in [1, 2, 10]. Note that in all experiments with exposure to neutrons [1, 2, 10], ions [3, 4] and superthreshold energy electrons [6] the bursts (i.e. the temperature oscillations) of only one type were found.

In the experiments [5] with irradiation by electrons of subthreshold energy, unlike the mentioned ones [1–4, 6, 10], two types of self-oscillations in the yield of particles from solid $CH_4$ were detected – with long and short periods. The first theoretical model of the self-oscillations which realized in solid $CH_4$ exposed to an electron beam of subthreshold energy was suggested in the study [24]. The formation of self-oscillations of temperature and concentration of radicals in an electron-irradiated methane film at low temperatures have been simulated and investigated experimentally. Calculations performed in [24] showed that, along with the temperature self-oscillations with a long period, which manifested themselves in a sharp and large rise in temperature, found earlier in neutron irradiated solid $CH_4$ [1, 2, 10], there are oscillations with smaller temperature fluctuations and a much shorter period. Concentration self-oscillations of two types of particles have been found and investigated, namely, H atoms and CH$_3$ radicals formed during the irradiation of methane by subthreshold energy electrons. The found two types of oscillations are a periodic change of temperature and concentration of these particles with irradiation time. It has been demonstrated that the temporal dynamics of the delayed desorption repeats the dynamics of self-oscillations in solid methane.

In order to further deepen the understanding of the phenomenon, the study of radiation effects induced by electron irradiation of the matrix isolated CH$_4$ was started [25–27]. Solid Ar was chosen as the matrix taking into account that the energy transfer to the dopant in Ar matrix is quite efficient due to overlapping of the Ar free exciton band (12.1 eV [28]) with the absorption band of CH$_4$ molecule [29]. Moreover, the most intense intrinsic emission of the matrix – self-trapped excitons STE at 9.8 eV also falls into the absorption band of methane and can efficiently excite methane molecule stimulating its dissociation *via* photon-induced processes. The Ar matrix is transparent to the STE emission because the STE states lie below the intrinsic absorption threshold of the matrix, which ensures uniform excitation of dopant centers throughout the entire volume of the matrix. Therefore, although low-energy electrons directly excite a thin surface layer of the sample, the rest of the Ar matrix (more than 99%) is excited by excitons and photons, and this case should be considered as "intrinsic photolysis".

Here we present an outgrowth of this research promoting the developed methods of correlated in real time emission spectroscopy of photons and particles. Nonstationary luminescence (NsL) [30] and nonstationary desorption (NsD) [31] methods, based on monitoring of selected emissions upon external heating of the doped matrices under an electron beam were applied. These measurements in combination with the detection of thermally stimulated exoelectron emission (TSEE) and optical emission spectroscopy (OES) made it possible to determine the contribution of the reactions of charged and neutral species. Unlike much of the research done with absorption spectroscopy, this approach is based on highly sensitive methods of emission spectroscopy, which include the study upon nonstationary conditions and registration of relaxation emissions. Such an approach enabled us to monitor spectroscopically by photon emissions different radiolysis products and trace simultaneously particle emission. New results in the study of radiation effects in matrix-isolated $CH_4$ are discussed with a focus on particle emission and phenomenon of self-oscillations.

## 2 Experimental

The developed approach for investigation of the radiation effects in cryogenic ices has previously been described in Sec.7 of [32] therefore only new and important details relevant to the present study are given here. The study was carried out in the range of $CH_4$ concentration from 1 to 10%. Mixture of Ar and $CH_4$ of defined concentration was performed in the stainless steel gas-handling system which was heated and degassed before each experiment. We used Ar gas (99.998%) and $CH_4$ gas (99.97%) without further purification. Films of solid Ar doped with methane were grown by deposition of a certain amount of premixed gas of room temperature onto a cooled to liquid Helium (LHe) temperature oxygen-free Cu substrate mounted in a high-vacuum chamber with a base pressure of $<10^{-7}$ torr. Cooling was accomplished with a LHe cryostat. 25 μm thick films used in these experiments were quite transparent and did not require annealing. The specified mode of sample heating was provided by a temperature controller. The sample temperature was monitored with a Si sensor.

The irradiation was performed in dc regime with subthreshold energy electrons ($E_e$<1.7 keV) to avoid the knock-on defect formation and sputtering. In these experiments the electron beam energy $E_e$ was set to 1.5 keV with the current density of 2.5 mA cm$^{-2}$. The beam covered the icy film with an area of 1 cm$^2$. The sample heating when turning on the electron beam did not exceed 0.4 K. Cathodoluminescence (CL) spectra were registered simultaneously in the VUV range and in the visible one. The dose deposited by irradiation was determined from the exposure time for a constant beam intensity. Particle desorption under beam was monitored with an ionization detector (a Bayard-Alpert gauge) throughout the entire experiment. On completing irradiation an "afteremission" current was detected due to the draining of excess charge. The measurements of relaxation emissions were started when the "afteremission" current had decayed to essentially zero. Three series of experiments were performed with different irradiation and heating modes: (i) continuous irradiation at low temperature with accumulation of radiolysis products, (ii) external heating of the pre-irradiated films with measurements of relaxation emissions (specifically thermally stimulated exoelectron emission TSEE) and (iii) external heating of the pre-irradiated films under beam. The heating mode used in cases (ii) and (iii) was identical – heating with a constant rate of 6 K min$^{-1}$. Released from shallow traps electrons being promoted to the conduction band either neutralize positively charged centers or escape from the film yielding TSEE current. Stimulated currents were detected with an electrode kept at a small positive potential $V_F$=+9 V and connected to the current amplifier. TSEE output reflects the distribution of shallow electron traps (defects of the structure) in the sample and provides information about the levels of traps in the band gap of the matrix. Owing to the high mobility of electrons in solid Ar [33], TSEE monitoring makes it possible to obtain information on relaxation processes not only near the surface, but also in the bulk of the matrix. Mode (iii) is a measurement of the photon yield at selected wavelengths from species of interest under non-stationary excitation conditions, namely, when a pre-irradiated sample is heated under an electron beam (that is a measurement of NsL and simultaneously NsD). The idea behind these pump-probe measurements is as follows. The first stage is the generation of charged and reactive species; the second stage corresponds probing the species under study with a low density beam (to diminish generation of new species) on heating. Wherein, the electrons released from traps during heating recombine with positively charged particles in neutralization reactions, which contribute to the NsL spectra of the corresponding species. Measured at the selected wavelength NsL curves are, in fact, a sum of spontaneous CL excited by a lower density electron beam and luminescence stimulated by gradual heating which, in turn, contains the contributions of charged and neutral species. In order to confidently single out the contribution of charged species and neutralization reactions, the NsL curves were compared with the curves of thermally stimulated exoelectron emission TSEE. It is clear that the contribution of luminescence stimulated by an electron release from shallow traps followed by their recombination with positively charged species reflects the trap energy distribution in the band gap and has a nonmonotonic character. Along with NsL curves, curves of nonstationary desorption NsD were recorded in a correlated manner.

Note that all measurements, both while the beam is on, and after its shutdown and subsequent heating, were performed in the dynamic pumping mode. Pumping out was carried out by LHe cryogenic pump and magnetic discharge pump. Time correlation of measurements of all parameters during the entire experiment was provided by a dedicated software.

**3 Result and discussion**

Monitoring of the electron-induced desorption from films of solid Ar doped with 10% of $CH_4$ via a pressure registration in the experimental chamber revealed the delayed explosive desorption of particles. The irradiation with a 1.5 keV electrons at electron current density j = 2.5 mA $cm^{-2}$ resulted in a pressure rise by an order of magnitude after long irradiation (about an hour) as shown in Figure 1. The rise in the delayed desorption yield occurred much faster than its decay.

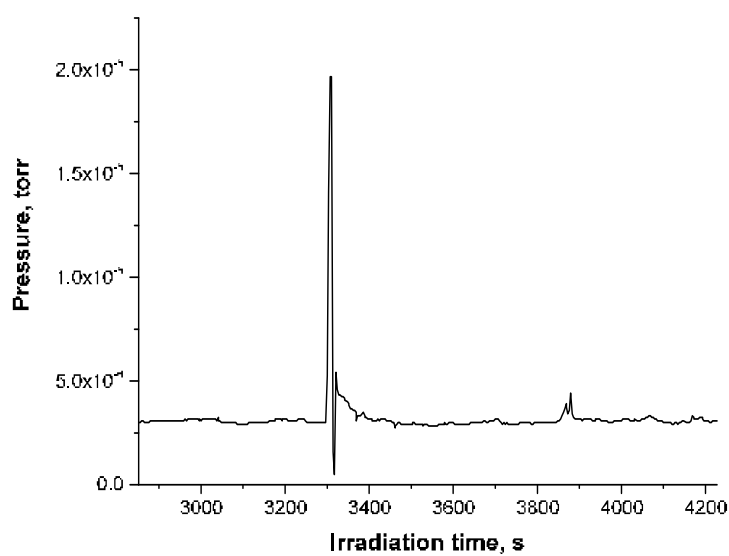

Fig.1 Delayed desorption from Ar matrix doped with 10% $CH_4$. Irradiation was performed at LHe bath temperature provided by a cryostat with electrons of 1.5 eV and j = 2.5 mA cm$^{-2}$.

And, what is more, we observed two bursts of the desorption which are a manifestation of self-oscillations in an irradiated Ar matrix containing methane. When discussing this effect, it should be taken into account that the measurements were carried out in the dynamic pumping mode. This means that in a closed chamber the effect would be much more pronounced. Note that no desorption bursts were found in a pure Ar matrix irradiated under the same conditions. The delay time $\tau_1$ for the first burst was $\tau_1$=3290 s. The delay time $\tau_2$ for the second burst is significantly shorter (it was about $\tau_2$ = 500 s after the end of the first burst) because a part of the radiolysis products remained in the sample after the first burst but the intensity of the second burst turned out to be much less, since most of the sample evaporated during the first burst. Both of these bursts have some structure shown in Figures 2a and b. So in this case we observe long-period bursts with a limited number of short-period oscillations. The structure components of both bursts are separated by 10 s.

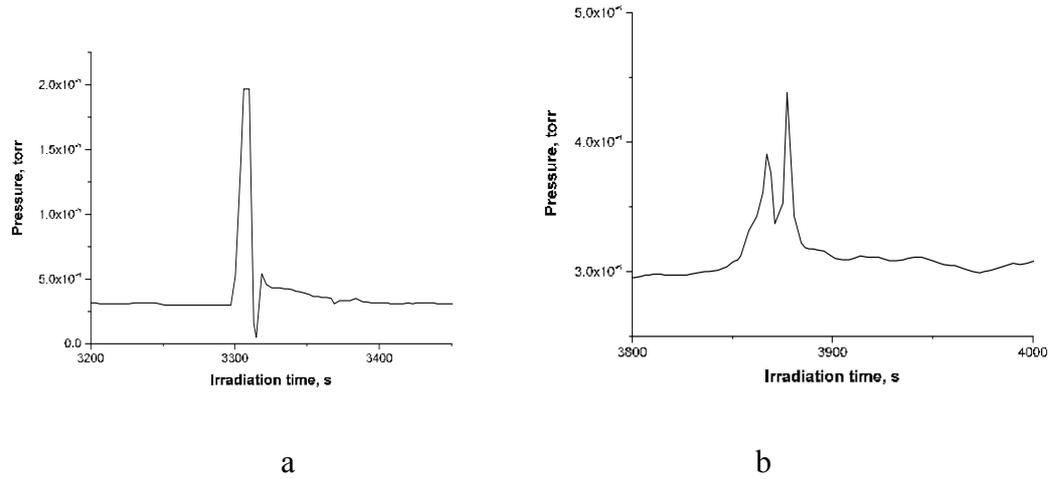

                    a                                b

Fig.2 Structure of the first (a) and second (b) bursts of particles from solid Ar doped with 10% $CH_4$.

It is of interest to compare the obtained result with that for Ar matrix containing 1% $CH_4$. The sample thickness and conditions of irradiation were identical with those kept for the Ar matrix doped with 10% $CH_4$. Bursts of delayed desorption, which were also observed at an order of magnitude lower methane concentration, are shown in Fig. 3 by the data presented in Ref. [25]. In this case, two desorption bursts were also found with a high intensity of the first peak and a reduced intensity of the second. The delay time $\tau_1$ for the first burst $\tau_1 = 3540$ s is longer than that for the Ar matrix doped with 10 % $CH_4$ as well as the delay time $\tau_2$ for the second burst was about $\tau_2 = 1200$ s after the end of the first one.

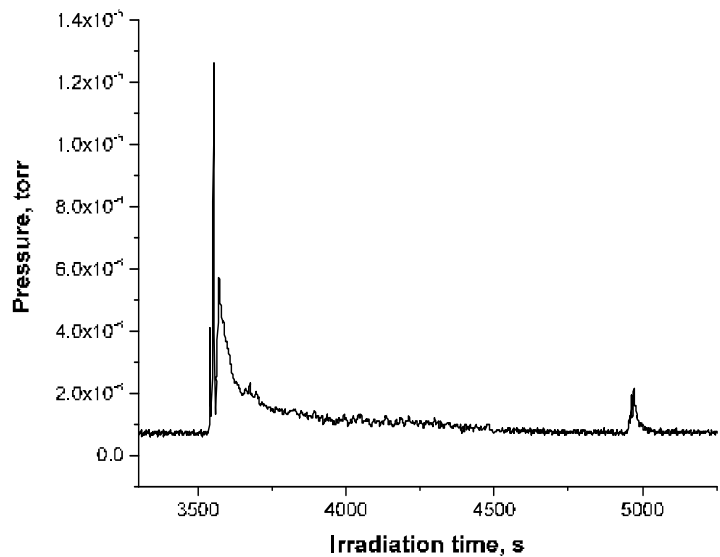

Fig. 3. Delayed desorption from Ar matrix doped with 1% $CH_4$. Irradiation was performed at LHe bath temperature provided by a cryostat with electrons of 1.5 keV and j = 2.5 mA cm$^{-2}$.

The first burst of particles desorption is composed of three peaks spaced apart by more than 10 seconds each. The peaks of the second burst are of smaller amplitude. They are rising with a period of about 7 sec and then fade out. So for the methane concentration 1% we observe also long period bursts with a limited number of short period self-oscillations shown separately in Figures 4 a and b.

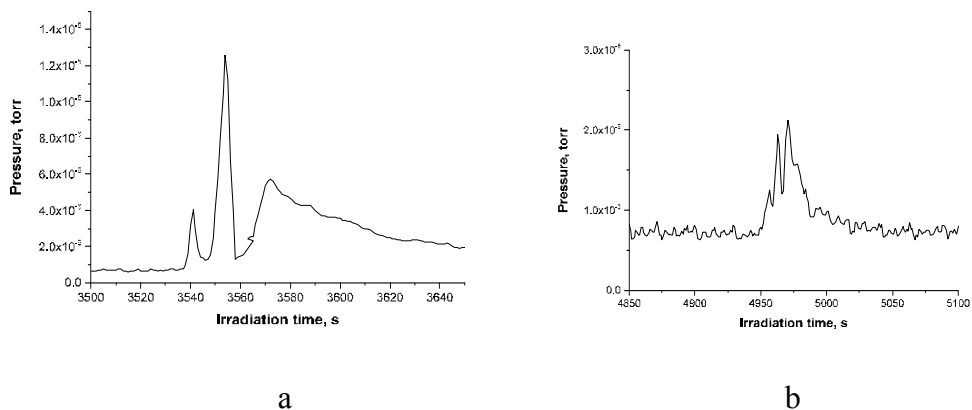

a b

Fig. 4 Structure of the first (a) and second (b) bursts of particles from solid Ar doped with 1% $CH_4$.

Delay times $\tau_1$ of bursts detected from Ar matrices doped with 1% and 10% CH$_4$ do not differ much and are close to that observed in pure methane $\tau$ = 3570 s at slightly different irradiation conditions, viz. – a 1 keV electron beam of a 3 mA cm$^{-2}$ density [5]. This suggests at least partial formation of CH$_4$ clusters in the Ar matrix. Short-period self-oscillations are not so extended as in pure methane and have only limited number of periods.

As it was mentioned, electronically excited states of CH$_4$ are of dissociative nature and one can expect to see in the CL spectra products of radiolysis of dopant. Figure 1 shows as an example, the typical CL spectrum of an Ar matrix doped with 1% CH$_4$, since at higher methane concentrations the CL intensity is strongly suppressed

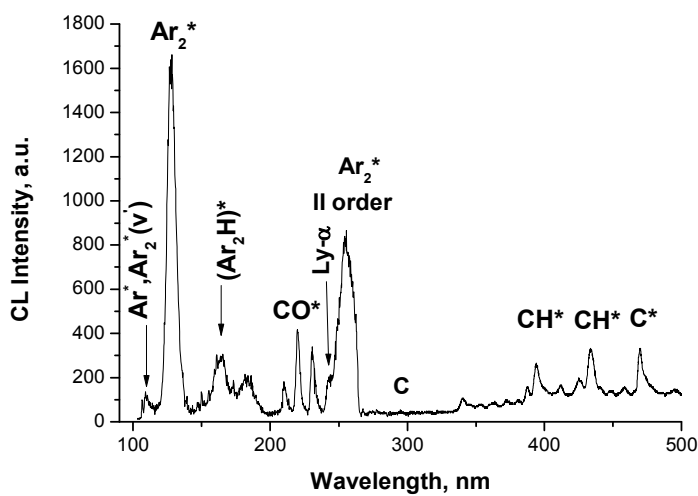

Fig. 5 CL spectrum of solid Ar doped with 1% CH$_4$ taken upon irradiation at 4.6 K. The CL spectrum, identical to those previously published [26, 27], consists of the emissions of the matrix and the dopant fragments. The most intense feature of this spectrum at 127 nm is the well-known emission of molecular-type STE – Ar$_2^*$, corresponding to the transitions from $^{1,3}\Sigma_u^+$ states to a repulsive part of the ground state $^1\Sigma_g^+$ [28]. This band was also detected in the second order. At the short-wavelength edge of this band, recorded in the second order, there is a feature – the Lyman-α line, which belongs to the excited hydrogen atoms that have desorbed into the gas phase. Mechanisms of the excited H atoms' formation and their desorption from Ar matrices, doped with $CH_4$, were discussed recently [34]. A weak unresolved feature at about of 109 nm is associated with electronically excited $Ar^*$ atoms and unrelaxed $Ar_2^*$ molecules desorbing in excited states. The spectrum also contains a broad band at 184 nm unidentified at present and impurity bands of CO (the Cameron system $a\,^3\Pi \rightarrow X^1\Sigma^+$). The most interesting feature in the VUV range related to the present study is the excimer emission band near 166 nm, which corresponds to the emission of the $Ar_2H^*$ center at the transition to the repulsive part of the ground state potential curve [35]. This center is one of the main products of radiolysis – H atom in the volume of the matrix. Main channels of $CH_4$ degradation (see e.g. [36] where results of numerous studies were compiled) are:

$$CH_4 + h\nu \rightarrow CH_3 + H \qquad (1)$$

$$CH_4 + h\nu \rightarrow CH_2 + H_2 \qquad (2)$$

$$CH_4 + h\nu \rightarrow CH_2 + 2H \qquad (3)$$

$$CH_4 + h\nu \rightarrow CH + H + H_2 \qquad (4)$$

Hydrogen atoms appear at methane molecule dissociation in reaction (1), (3) and (4). When excited by the STE excitons, the branching ratio BR for the first reaction is BR=0.5 according to Ref. [37] in which an analysis of energy-dependent branching ratios, the so called breakdown curves, was presented. BRs in reactions (3) and (4) are less than 0.2. In the visible range we observed emission bands of other radiolysis products – CH radical and C atom. CH radicals were registered by the emission bands at 432 nm (the $A^2\Delta \rightarrow X^2\Pi$ transition) and 387 nm (the $B^2\Sigma^- \rightarrow X^2\Pi$ transition). C atoms were recorded by the emission lines at 470 nm (the $^1S \rightarrow\,^3P$ transition) and 295 nm (the $^5S^0 \rightarrow\,^3P$ transition). The appearance of C emission indicates the implementation of channel (5), for which the threshold wavelength of $CH_4$ dissociation is 151.1 nm according to [39].

$$CH_4 + h\nu \rightarrow C + 2H_2 \qquad (5)$$

Another way for the formation of C atoms is the successive dissociation of secondary radiolysis products: $CH_3$, $CH_2$, and CH.

Let us discuss a dynamics of H atoms and CH radicals at nonstationary conditions, i. e. monitored using mode (iii). Externally induced heating of sample results in various consequences. Thermally stimulated release of electrons from deeper and deeper traps contributes to the recombination of positively charged centers and the intensification of the emission bands of NsL of the corresponding neutralization products. This contribution can be identified by comparing the NsL curves with the TSEE yield. On the other hand, such heating stimulates thermally activated processes, in particular, the diffusion of radiolysis products and their subsequent reactions. CH radicals appear in reaction (4) with the BR of about 0.1. A distinct intensity of the CH emission (see Fig. 5) indicates the need to take into account also secondary reactions involving $CH_3$ and $CH_2$ fragments. An analysis of the photolysis channels performed in [27], taking into account secondary reactions showed that the main source of CH fragment formation is $CH_3^+/CH_3^*$ photolysis. In other words, the CH radical can be considered as a signature of the $CH_3^*$ species. Note, that the CH emission (the $A^2\Delta \to X^2\Pi$ transition) was used to visualize the distribution of $CH_3$ radicals in methane-air flames [38]. In our study we used the same emission band at 432 nm (the $A^2\Delta \to X^2\Pi$ transition) which was more intense to monitor NsL of CH radical. H atoms in the bulk of matrix were traced by the $Ar_2H^*$ emission at 166 nm. Figure 6 presents their behavior upon externally induced heating under an electron beam of pre-irradiated Ar matrix doped with 10% $CH_4$. The NsL curves of both species have a nonmonotonic character as can be seen from the Figure. On the one hand, the features of the temperature behavior of the NsL CH radical correlate with the yield of TSEE, which indicates their appearance in the neutralization reaction of $CH_3^+$. The emission maxima of NsL CH in the region of 10 – 15 K and about 35 K are close to the maxima recorded in the TSEE yield.

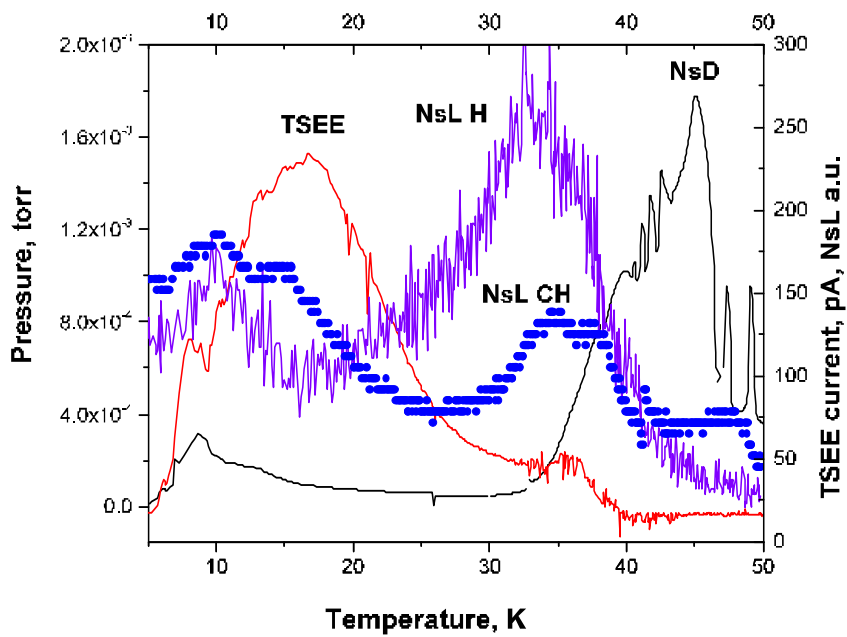

Fig.6 NsL curves of H atoms and CH radicals measured in a correlated way with the NsD yield from Ar matrix doped with 10% $CH_4$. The measurements were performed upon externally induced heating of a pre-irradiated sample under an electron beam. The same Figure shows the TSEE yield.

On the other hand, it is known that the characteristic temperature at which diffusion of ordinary particles in an Ar matrix first becomes noticeable is about 35 K [40], which gives grounds to attribute the decrease in the NsL intensity of the CH radical, which reflects the behavior of $CH_3$, to the recombination of diffusing $CH_3$ and the formation of $C_2H_6$ species with the release of energy spent on desorption. This scenario corresponds to the antibate behavior of the NsL curve for CH radicals and the NsD curve. Note that $CH_3$ radicals registered by FTIR absorption on 121.6 nm photolysis of methane in Ar matrix remained still stable at 14 K [41]. Formation of $C_2H_6$ species was reported in Ne matrix doped with 1% $CH_4$ on photolysis with 121.6 nm flux [39, 42]. H atoms which produced in reactions (1), (3) and (4) appeared to be quite "hot", i.e. have significant excess of kinetic energy. In channels (3) and (4) it exceeds 1 eV, and in channel (1) 3.1 eV [37], which contributes to the diffusion of H atoms over long distances until thermal equilibrium is established and they can recombine. It is established that the H atoms occupy two types of sites in Ar lattice – octahedral interstitial site and substitutional one [43, 44]. The atoms occupying the interstitial sites are released at moderate heating, while the atoms trapped in the substitutional sites remain trapped at low temperatures [45]. A decrease in the NsL intensity for H atoms, observed in the temperature range 32–45 K, is associated with thermally stimulated recombination of H atoms and formation of $H_2$ molecule. In this temperature range, the NsL H and NsD curves have an antibate character, which is thought to be due to the contribution of the recombination of H atoms with energy release to the stimulation of desorption. Such a behavior of the NsL H curve is characteristic only for samples with a high methane content of 10%, which indicates a significant role of the recombination of hydrogen atoms in the delayed explosive desorption of pure solid methane. The oscillations of the NsD, growing against the background of the peak of desorption, are of special attention. The characteristic time $\tau_{osc}$ of these oscillations is about $\tau_{osc} = 7$ s. We also observed oscillations on the recession of the peak of desorption with longer $\tau_{osc}$ of about 20 s. This phenomenon can be interpreted as thermo-concentration self-oscillations [24]. They can be initiated in two ways – by external heating of the sample to release the stored chemical energy in recombination reactions, or by spontaneous release of the stored energy after reaching critical concentrations of radicals. This phenomenon is similar to that observed in pure solid methane when irradiated with an electron beam [5, 24]. Long-period thermo-concentration self-oscillations were also observed upon neutron irradiation of solid methane at low temperature and in a course of externally induced heating [1].

In Figure 7 NsL curves of H atoms and CH radicals measured in a correlated way with the NsD yield from Ar matrix doped with 1% $CH_4$ are shown for comparison.

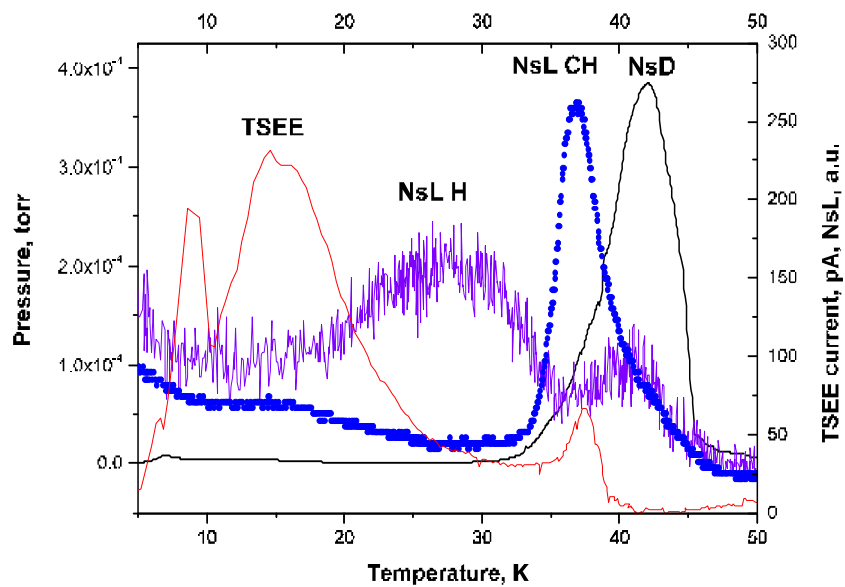

Fig. 7 NsL and NsD observed from 1% $CH_4$ in Ar matrix when heated under the beam in comparison with the TSEE yield.

The dynamics of all emissions is nearly identical to that obtained in our previous study published in [27]. As it can be seen there is no correlation in behavior of NsL emissions of H atoms and CH radicals and antibate behavior of NsD and NsL was observed only for CH radical, that is for $CH_3$ species for lightly doped matrices. The dynamics of NsL H follows the dynamics of NsD in this case. Antibate behavior of the NsL H and NsL CH/$CH_3$ may occur because of most of these fragments appear in different reactions at low concentrations. Another reason may be associated with the movement of hot hydrogen atoms over long distances, which affects the conditions for their recombination. It should be noted that NsD of methane-containing Ar matrices includes NsD of the matrix itself and desorption caused by reactions of radiolysis products. A significant difference in NsD matrices, doped with 1 and 10% $CH_4$, demonstrates the decisive effect of doping. The contribution from electronically stimulated desorption is small due to the low intensity of the beam used at the probing stage. Future studies are invited to systematically investigate the influence of matrix composition and a wider row of radiolysis products on the ability of a cryogenic environment to store chemical energy and stimulate self-oscillation processes.

**4 Summary**

The phenomenon of delayed explosive desorption of particles stimulated by irradiation with electrons was studied for matrix-isolated methane in the extended concentration range from 1% to 10%. Solid Ar was chosen as a matrix because of high efficiency of the energy transfer to the dopant due to overlapping of the Ar free and self-trapped exciton bands with the absorption band of $CH_4$. The study was performed using high sensitivity emission spectroscopy methods – CL, measurements of relaxation emissions, particularly TSEE and registration of spectrally resolved nonstationary emissions of photons (NsL) and particles (NsD). Dynamics of main products of methane fragmentation: H atoms and $CH_3$ radicals was monitored upon external heating under the beam. An information on behavior of $CH_3$ radicals was obtained taking into account that the CH radical can be considered as a signature of the $CH_3$ species [27]. We found two bursts of long-period self-oscillations together with short-period ones with a limited number of periods upon stationary irradiation of all methane-doped Ar matrices at low temperatures. The experiments with external heating under the beam enlighted the decisive role of H atoms' and $CH_3$ radicals' recombination reactions in stimulation of the delayed desorption from heavily doped Ar matrices.

**Acknowledgements** The authors cordially thank colleagues Vladimir Sugakov, Giovanni Strazzulla, Oleg Kirichek, Robert Kołos, Claudine Crepin-Gilbert and Hermann Rothard for stimulating discussions.